\documentclass[a4paper]{article}

\usepackage{INTERSPEECH2020}
\usepackage{mathrsfs}
\usepackage{diagbox}
\usepackage{multirow}
\usepackage{subfigure}
\usepackage{threeparttable}
\usepackage{algorithm}
\usepackage{algorithmic}

\title{TSNAT: Two-Step Non-Autoregressvie Transformer Models \\ for Speech Recognition}
\name{Zhengkun Tian$^{1,2}$, Jiangyan Yi$^{1}$, Jianhua Tao$^{1,2,3}$, Ye Bai$^{1,2}$, Shuai Zhang$^{1,2}$, \\ Zhengqi Wen$^{1}$, Xuefei Liu$^{1}$}
\address{
	$^1$NLPR, Institute of Automation, Chinese Academy of Sciences, Beijing, China\\
	$^2$School of Artificial Intelligence, University of Chinese Academy of Sciences, Beijing, China\\
	$^3$ CAS Center for Excellence in Brain Science and Intelligence Technology, Beijing, China}
\email{\{zhengkun.tian, jiangyan.yi, jhtao, ye.bai, shuai.zhang, zqwen, xuefei.liu\}@nlpr.ia.ac.cn}

\begin{document}

\maketitle
\begin{abstract}
The autoregressive (AR) models, such as attention-based encoder-decoder models and RNN-Transducer, have achieved great success in speech recognition. They predict the output sequence conditioned on the previous tokens and acoustic encoded states, which is inefficient on GPUs. The non-autoregressive (NAR) models can get rid of the temporal dependency between the output tokens and predict the entire output tokens in at least one step. However, the NAR model still faces two major problems. On the one hand, there is still a great gap in performance between the NAR models and the advanced AR models. On the other hand, it's difficult for most of the NAR models to train and converge. To address these two problems, we propose a new model named the two-step non-autoregressive transformer(TSNAT), which improves the performance and accelerating the convergence of the NAR model by learning prior knowledge from a parameters-sharing AR model. Furthermore, we introduce the two-stage method into the inference process, which improves the model performance greatly. All the experiments are conducted on a public Chinese mandarin dataset ASIEHLL-1. The results show that the TSNAT can achieve a competitive performance with the AR model and outperform many complicated NAR models.

\end{abstract}
\noindent\textbf{Index Terms}: Autoregressive, Non-Autoregressive, Transformer, Two-Step, Speech Recognition

\section{Introduction}

End-to-end models have achieved great success in speech recognition, especially attention-based encoder-decoder models \cite{chorowski2015attention, chan2016listen, kim2017joint, dong2018speech} and transducer-based models \cite{graves2012sequence, he2019streaming, Tian2019, zhang2020transformer, yeh2019transformer}. Most of these end-to-end models generate the target sequence in an autoregressive fashion, which predicts the next token conditioned on the previously generated tokens and the acoustic encoded sequence. The autoregressive characteristic makes the inference process must be carried out step by step, which cannot be implemented in parallel and results in a large latency. By contrast, the non-autoregressive model (NAR) can get rid of the temporal dependency and directly generate the target sequences based on the encoded acoustic states in at least one step.

Although the Non-autoregressive models can perform inference very efficiently, it still faces two significant problems. On the one hand, there is still a great gap in performance between the advanced autoregressive (AR) model and the non-autoregressive model. Chen et.al proposed two kinds of iterative inference methods to alleviate the problem \cite{chen2019non}. Too many iterations have a negative impact on the speed of inference. Besides, some researchers also utilized a CTC Model to generate the preliminary predictions and then correct the previous prediction by a non-autoregressive decoder \cite{higuchi2020mask, higuchi2020improved, song2020non}. However, it will introduce some new problems. It is hard to eliminate accumulated error caused by the CTC Models and carry out the frame-wise operation of the CTC module in parallel. On the other hand, it is difficult for most of the non-autoregressive transformer models to train and converge. To our knowledge, there are three major ways to alleviate this problem. Firstly, some works try to introduce an auxiliary CTC loss to accelerate the training and convergence \cite{higuchi2020improved, Tian2020}. Secondly, more iterations in the training process also bring the improvement on the performance \cite{chen2019non, Bai2020}, which is very simple and straightforward. Thirdly, instead of learning from a sequence partially or fully filled with \texttt{<MASK>} \cite{chen2019non, Bai2020}, some works try to improve training efficiency by providing a preliminary sequence with the information of target sequence to the non-autoregressive transformer decoder \cite{higuchi2020mask, higuchi2020improved, song2020non, Tian2020}. These methods are either time-consuming or difficult to implement.

The traditional NAR model predicts the output tokens conditional on the acoustic encoded states and a sequence that is completely empty and does not contain any prior knowledge. In order to make the training match the inference, the NAR model tries to learn from the sequence fully filled with \texttt{<MASK>} in the training process, which hinders the improvement of training speed and performance. To address this problem, inspired by the dual-mode ASR (streaming and non-streaming) \cite{yu2021dual} and two-pass end-to-end models \cite{sainath2019two, hu2020deliberation, tian2020one}, we propose a model named two-step non-autoregressive transformer (TSNAT). The TSNAT utilizes an AR model to accelerate the NAR model convergence and improve the performance. Our model consists of an optional frond-end block, an acoustic encoder, and a dual-mode transformer decoder. The dual-mode transformer decoder can model the context in both an autoregressive and a non-autoregressive way. Different from the traditional method that transfer knowledge into the NAR model from a pre-trained AR model \cite{ren2019fastspeech}, we train an AR model and a NAR model from scratch simultaneously. The NAR model can learn some linguistic prior knowledge from the AR model depending on the dual-mode transformer decoder. During the inference, our model can perform two-step decoding without depending on the external attention decoder. All the experiments are conducted on a public Chinese mandarin dataset ASIEHLL-1. The results show that the TSNAT can achieve a competitive performance with the AR model and outperform other NAR models.

\begin{figure*}[t]
    \centering
    \subfigure[The Structure of Two-Step Non-Autoregressive Transformer]{
        \centering
        \label{fig:enc_and_infer}
        \includegraphics[width=0.6\linewidth]{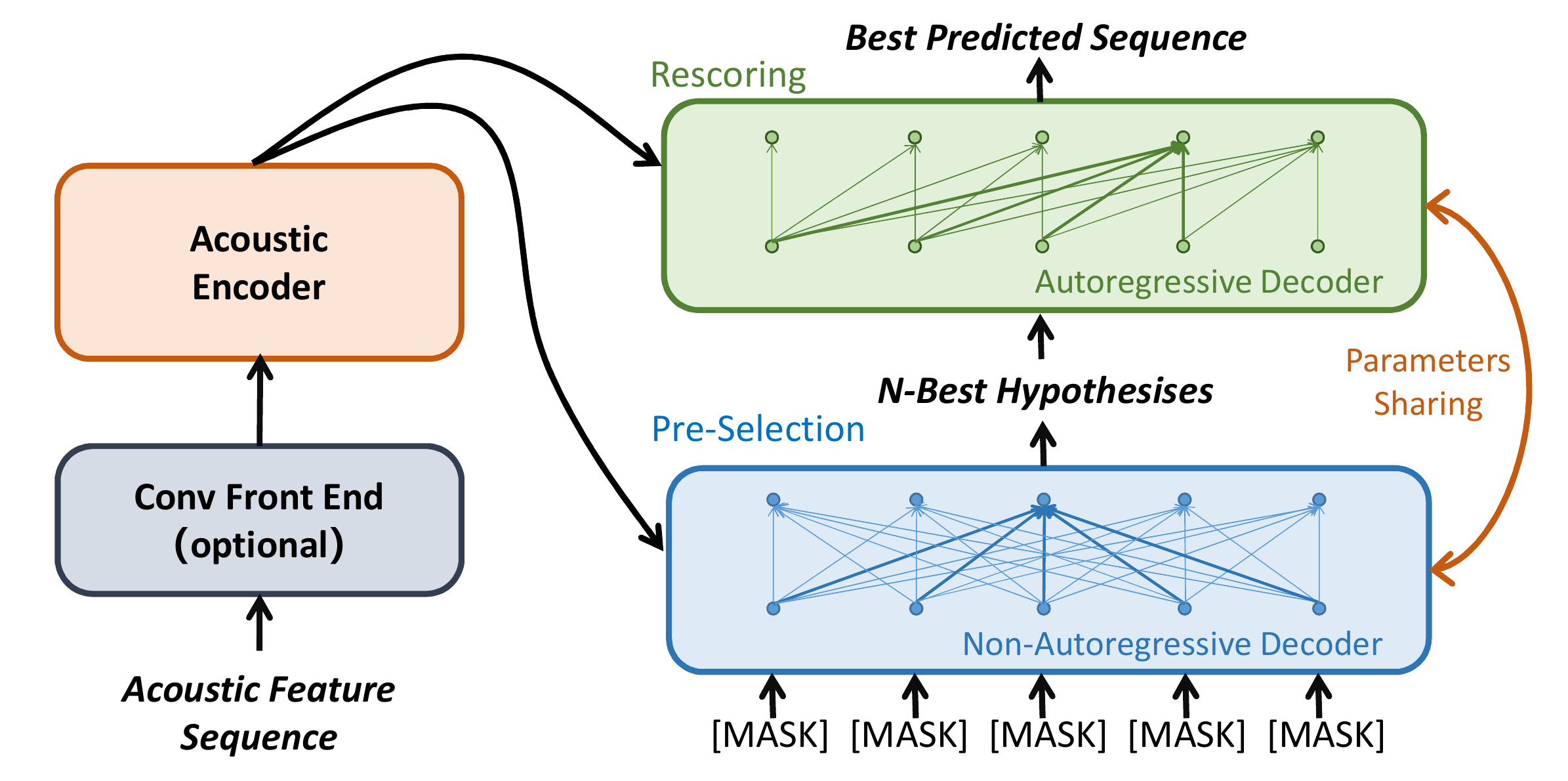}}
    \subfigure[The First-Step Inference Graph]{
        \centering
        \label{fig:decoder}
        \includegraphics[width=0.35\linewidth]{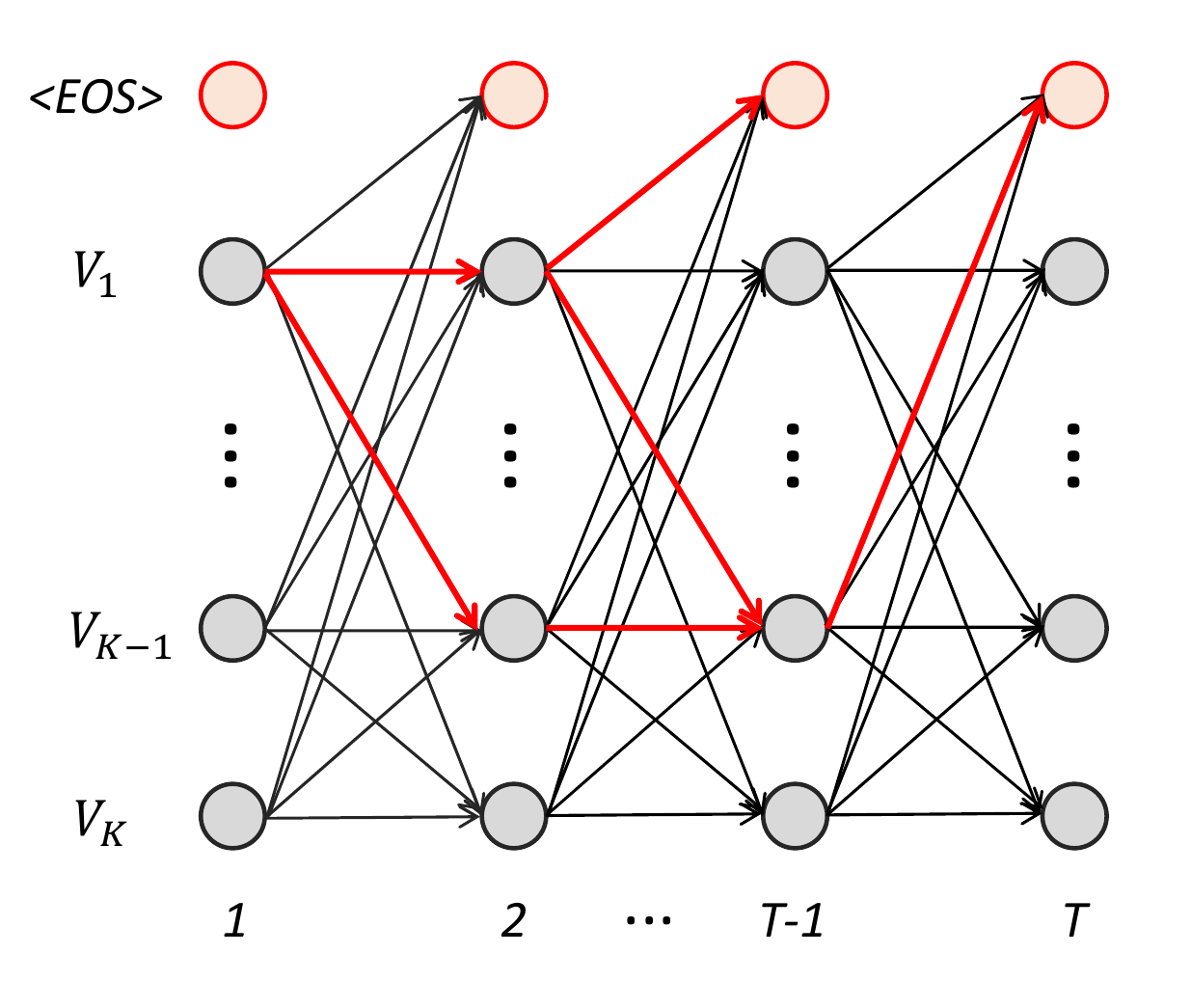}}
    \vspace{-10pt}
    \caption{(a) illustrates the structure of our proposed two-step non-autoregressive transformer (TSNAT) and the inference process. The TSNAT model consists of an optional front-end block, an acoustic encoder, and a dual-mode transformer decoder. The inference can be divided into two-step, pre-selection and rescoring. The dual-mode decoder first generates n-best hypothesizes in a non-autoregressive and then rescores these hypotheses in an autoregressive way. (b) illustrates an output probability graph of the non-autoregressive model. The first red circle of each column means the end-of-sentence token \texttt{<EOS>}.  The other black circle represents different tokens of vocabulary. Each path that starts with the blank circle and ends with a red circle in the graph represents a possible hypothesis.}
    \vspace{-10pt}
\end{figure*}

The remainder of this paper is organized as follows. Section 2 describes our proposed model. Section 3 presents our experimental setup and results. The conclusions will be given in Section 4.

\section{Two-Step Non-Autoregressive Transformer}
Our proposed two-step non-autoregressive transformer, as shown in Fig.1(a), consists of three components, an optional convolutional front end block, an acoustic encoder, and a dual-mode transformer decoder.
\vspace{-5pt}
\subsection{The Convolutional Front End and Acoustic Encoder}
\vspace{-5pt}
The convolutional front end block generally consists of two 2D-Convolution layers and an output project layer \cite{dong2018speech}. The convolution layer is mainly responsible for the processing and down-sampling of the low-level acoustic features. The last output project layer will project the processed acoustic feature into the dimension that the encoder required. The convolutional front end block also can be replaced with the linear project layers with splicing frame operation \cite{8682586}.

We utilize the transformer encoder \cite{vaswani2017attention, dong2018speech} as the acoustic encoder, which contains a positional embedding, $N$ repetitive multi-head self-attention layers (MHA), and feed-forward network layer (FFN). The sine-cosine positional embedding proposed by \cite{vaswani2017attention} is applied for all the experiments in this paper. Besides, the model also applies residual connection and layer normalization.

The multi-head self-attention layers depend on all contexts to compute the attention weights from different perspectives, which makes it powerful to model long-range temporal information.
\vspace{-5pt}
\begin{equation}
    \begin{split}
        \text{SA}_i(\bm{Q}, \bm{K}, \bm{V})&=\text{Softmax}((\bm{Q}_t)\bm{K}^{\top}/\sqrt{d_k})\bm{V}\\
        \text{MHA}(\bm{Q}, \bm{K}, \bm{V})&=\text{Concat}(SA_0, SA_1, ..., SA_H)
    \end{split}
\end{equation}
where $\bm{Q}$, $\bm{K}$ and $\bm{V}$ indicate the query, key, and value matrix respectively. $d_k$ is the last dimension of $\bm{K}$. The multi-head attention concatenates $H$ output vectors of the self-attention mechanism. Besides, both the input of self-attention and the output of multi-head attention should equip with an independent weight matrix respectively. The feed-forward network (FFN) contains two linear layers and a gated liner unit (GLU) \cite{dauphin2017language} activation function \cite{Tian2019}.
\vspace{-5pt}
\begin{equation}
    \text{FFN}(x)=\text{GLU}(xW_1+b_1)W_2+b_2
\end{equation}
where parameters $W_*$ and $b_*$ are learnable weight matrix and bias respectively.
\vspace{-5pt}
\subsection{The Dual-Mode Transformer Decoder}
\vspace{-5pt}
The major difference between the autoregressive model and the non-autoregressive model focus on the structure of the decoder. The AR models force themselves to learn the linguistic dependencies by blocking out the future information. This characteristic makes the AR model predict the output sequence step by step. The conditional probability $P_{AR}(Y|X)$ can be expressed as:
\vspace{-5pt}
\begin{equation}
    P_{AR}(Y|X)=P(y_1|X)\prod_{i=2}^{L}P(y_i|y_{<i},X)
\end{equation}
where $X$ is the acoustic encoded sequence with length $T$ and $y_i$ denotes the $i$-th token of predicted sequence with length $L$. However, the non-autoregressive models get rid of the dependencies on the previously predicted tokens. Each step in the inference process is independent of each other, which makes the inference step can be implemented in parallel and greatly improves the reasoning speed of the model. The conditional probability $P_{NAR}(Y|X)$ can be rewritten as:
\vspace{-5pt}
\begin{equation}
    P_{NAR}(Y|X)=\prod_{i=1}^{L}P(y_i|X)
\end{equation}

Both the AR decoder and the NAR decoder can apply the transformer as the basic structure. There are two significant differences between these two kinds of decoders. On the one hand, the AR transformer needs to apply the masking operation to the previous output to model the linguistic dependencies, but the NAR transformer needs to model the bidirectional dependencies between the output tokens without any masking operations. On the other hand, the AR decoder adopts all the tokens of vocabulary and some special tokens as the modeling units, while the NAR decoder only requires one \texttt{<MASK>} token. Their modeling units are not coincident.

Inspired by the dual-mode ASR \cite{yu2021dual}, which models the streaming ASR and non-streaming ASR by sharing an encoder, we propose a novel method that unifies the AR and NAR decoder into one dual-mode transformer decoder (DMTD). We assume that the AR decoder will accelerate the training and convergence of the NAR decoder. The linguistic dependencies learned by the AR decoder might be able to provide some prior knowledge for the NAR decoder, which will make the training of the NAR decoder more efficient than guessing based on the empty sequence. The self-attention of DMTD applies to the masking-optional operation. During training, the encoder forwards once, and the decoder forwards twice, once in autoregressive mode and once in non-autoregressive mode. During the AR forward, the DMTD will adopt the truth token sequences that begin with the begin-of-sentence token \texttt{<BOS>} as the input, and then mask future information of self-attention and calculate the cross-entropy (CE) loss $\mathcal{L}_{AR}$. For the NAR forward, the decoder will utilize a sequence filled with \texttt{<MASK>} as the input and calculate the CE loss $\mathcal{L}_{NAR}$. The final loss is equal to the weighted sum of $\mathcal{L}_{AR}$ and $\mathcal{L}_{NAR}$.
\begin{equation}
   \mathcal{L} = (1 - \alpha)\mathcal{L}_{NAR} + \alpha \mathcal{L}_{AR}
\end{equation}
where $\alpha$ dictates the weight of AR loss. The model is optimized by these two losses jointly.

\subsection{The Two Step Inference}
Most of the non-autoregressive models just select the token which has the highest probability at each position and concatenates them from left to right as the final predicted result \cite{chen2019non, Tian2020,Bai2020}. This inference method can be regarded as a greedy search. As shown in Fig.1(b), we consider the conditional probability matrix generated by the decoder as a graph, which contains numerous possible hypothesize. Each hypothesis starts with any blank circle in the first column and ends with a red circle (end-of-sentence token \texttt{<EOS>}). We could select a better hypothesis than the one predicted by greedy search by depending on the external language model or the second attention decoder as \cite{sainath2019two, tian2020one, hu2020deliberation}. Now that the dual-mode transformer decoder is able to be trained in the AR and NAR fashion simultaneously, the decoder can perform dual-mode inference naturally.

The inference process can be divided into two steps, pre-selection and rescoring. During the first pre-selection, the dual-mode decoder will generate the conditional probability matrix from a full-mask sequence in a NAR way and then select the $N$-best hypothesizes. In order to avoid that the model tends to choose shorter sentences, we utilize the length-normalized scores. The selection of $N$-best hypothesizes from the probability graph contains only simple addition operations, and does not take much time and computing resources. During the second step, the dual-mode decoder will insert begin-of-sentence token \texttt{<BOS>} into the head of the $N$-best hypothesizes. Then it utilizes the processed candidates as the input and predict them in an AR fashion. Due to the transformer can carry out efficient computing in parallel, the entire inference process can be finished in two step.


\section{Experiments and Results}
\vspace{-5pt}
\subsection{Dataset}
\vspace{-5pt}
In this work, all experiments are conducted on a public Mandarin speech corpus AISHELL-1\footnote{https://openslr.org/33/}.The training set contains about 150 hours of speech (120,098 utterances) recorded by 340 speakers. The development set contains about 20 hours (14,326 utterances) recorded by 40 speakers. And about 10 hours (7,176 utterances / 36109 seconds) of speech is used as the test set. The speakers of different sets are not overlapped.
\vspace{-5pt}
\subsection{Experimental Setup}
\vspace{-5pt}
For all experiments, we use 80-dimensional FBANK features computed on a 25ms window with a 10ms shift. We choose 4234 characters (including a padding symbol \texttt{<PAD>}, an unknown token \texttt{<UNK>}, a begin-of-sentence token \texttt{<BOS>}) and an end-of-sentence token \texttt{<EOS>} as modeling units.

Our model consists of 12 encoder blocks and 6 decoder blocks. There are 4 heads in multi-head attention. The 2D convolution front end utilizes two-layer time-axis CNN with ReLU activation, stride size 2, channels 384, and kernel size 3. Both the output size of the multi-head attention and the feed-forward layers are 384. The hidden size of the feed-forward layers is 768. We adopt an Adam optimizer with warmup steps 12000 and the learning rate scheduler reported in \cite{vaswani2017attention}. After 100 epochs, we average the parameters saved in the last 20 epochs. We also use the time mask and frequency mask method proposed in \cite{park2019specaugment} instead of speed perturbation.

We use the character error rate (CER) to evaluate the performance of different models. For evaluating the inference speed of different models, we decode utterances one by one to compute real-time factor (RTF) on the test set. The RTF is the time taken to decode one second of speech. All experiments are conducted on a GeForce GTX TITAN X 12G GPU.
\vspace{-5pt}
\subsection{Results}
\subsubsection{Comparison of The Model With Different AR Weights $\alpha$}
\vspace{-5pt}
This section compares the models with the different AR weights $\alpha$. When the weight $\alpha$ is equal to 0, the NAR model is equivalent to the one that adopts an empty sequence as the input without any prior knowledge. If $\alpha$ is equal to 1, the model will be completely transformed into an AR model. When the weight is between 0 and 1, the model can perform the two-step inference. For all one-step inference in this section (marked as \texttt{OneStep}), the NAR models ($\alpha \in [0, 1)$) apply greedy search and the AR ($\alpha=1.0$) models apply the beam search with width 10. The two-step inference of this section (marked as \texttt{TwoStep}) rescores the 10-best candidate sequences. As shown in Table 1, it's obvious that the hybrid AR and NAR training ($\alpha \in (0, 1)$) can improve the performance of the NAR model. We guess that the NAR decoder can get some prior knowledge from the AR decoder by sharing the parameters, which reduces the difficulty of training the NAR model from scratch. In the case of one-step inference, there is still a great performance gap between the AR model($\alpha=1.0$) and the NAR model ($\alpha \in [0, 1)$). However, the introduced two-step inference method can improve the performance of the NAR model greatly and achieve comparable results with the AR model.
\begin{table}[h]
    \caption{Comparison of the model with different AR weights $\alpha$ (CER \%).}
    \vspace{-5pt}
    \centering
    \label{tab:table1}
    \begin{tabular}{|c|c|c|c|c|c|c|c|}
        \hline
        \multicolumn{2}{|c|}{Weight $\alpha$}	 & 0.0 &  0.3  & 0.5 & 0.7 & 0.9 & 1.0 \\
        \hline
        \hline
        \multirow{2}{*}{Dev} & OneStep & 6.5 &  5.9 & \textbf{5.8} & \textbf{5.8} & 6.5 & 5.3 \\ \cline{2-8}
        & TwoStep & - & 5.6 & 5.5 & \textbf{5.4} & 5.6 & - \\
        \hline
        \hline
        \multirow{2}{*}{Test} & OneStep & 7.2 & 6.5 & 6.5 & \textbf{6.4} & 6.5 & 6.0 \\ \cline{2-8}
        & TwoStep & - & 6.2 & 6.1 & \textbf{6.0} & 6.2 & - \\
        \hline
    \end{tabular}
    \vspace{-10pt}
\end{table}
\vspace{-10pt}

\subsubsection{The influence of the $N$-best Sequences on The Model Performance}
This section pays attention to the inference of the $N$-best sequence on the NAR model performance. We select the different number of candidate sequences. Under the condition that $N$ is equal to 1, the two-step inference makes no sense. Therefore, we replace it with the result of the greedy search. The results in Table 2 indicate that the more candidate sequences, the better performance it archives, which is consistent with the results previously reported \cite{tian2020one}. When the number of candidate sequences is greater than 10, the performance tends to be stable. However, with the increase of $N$, the model requires more computing resources, which leads to an increase in latency and real-time-factor (RTF). Taken together, we will select the 10-best candidates for the second-step rescoring by balancing the model performance and the inference speed.

\begin{table}[h]
    \caption{Comparison of the influence of the $N$-best sequences on the model performance (CER \%).}
    \vspace{-5pt}
    \centering
    \label{tab:table2}
    \begin{tabular}{|c|c|c|c|c|c|c|}
        \hline
        $N$	 & 1 &  5  & 10 &  20 & 50 \\
        \hline
        \hline
        Dev & 5.8 &  5.4 & 5.4 & 5.4 & \textbf{5.3} \\
        \hline
        Test & 6.4 & 6.0 & \textbf{5.9} & \textbf{5.9} & \textbf{5.9} \\
        \hline
        RTF & 0.0054 & 0.0167 & 0.0173 & 0.0181 & 0.0220 \\
        \hline
    \end{tabular}
    \vspace{-10pt}
\end{table}
\vspace{-10pt}
\subsubsection{Comparison with Other Non-Autoregressive Models}
To further study the upper limit of model performance, we readjust the parameter configuration of the model and named three models of different sizes, named \texttt{TSNAT-Small}, \texttt{TSNAT-Middle} and \texttt{TSNAT-Big}. The three models have the same depth (12 encoder blocks and 6 decoder blocks) and are trained under the same conditions. Their differences focus on the dimensions of the model ($d_{model}$) and the hidden size of the feed-forward network ($d_{ff}$), as shown in Table 3.
\begin{table}[h]
    \caption{The Parameter Configuration of The Model.}
    \vspace{-5pt}
    \centering
    \label{tab:table3}
    \begin{tabular}{|c|c|c|c|}
        \hline
        Mode & TSNAT-Small &  TSNAT-Middle  & TSNAT-Big \\
        \hline
        \hline
        $d_{model}$ & 384 & 512 & 512 \\
        $d_{ff}$ & 384 & 512 & 512 \\
        \hline
        \hline
        \#Params & 34M & 59M & 87M \\
        \hline
    \end{tabular}
    \vspace{-10pt}
\end{table}
\vspace{-10pt}
\begin{table}[h]
    \caption{Comparison with other non-autoregressive models (CER \%). All models in this table use SpecAugment to improve the performance.}
    \vspace{-5pt}
    \centering
    \label{tab:table4}
    \begin{threeparttable}
        \begin{tabular}{|l|c|c|c|c|}
            \hline
            Model & LM & Dev & Test & RTF \\
            \hline\hline
            A-FMLM(K=1) \cite{chen2019non} & w/o & 6.2 & 6.7 & - \\
            \hline
            Insertion-NAT \cite{fujita2020insertion}  & w/o & 6.1 & 6.7 & - \\
            \hline
            LASO-big \cite{Bai2020} $\lozenge$ & w/o & 5.8 & 6.4 & - \\
            \hline
            CASS-NAT \cite{fan2020cass} $\lozenge$  & w & 5.3 & 5.8 & - \\
            \hline
            CTC-enhanced NAR \cite{song2020non} $\lozenge$ & w/o & 5.3 & 5.9 & - \\
            \hline
            ST-NAT \cite{Tian2020} & w/o & 6.9 & 7.7 & 0.0056 \\
            ST-NAT \cite{Tian2020}  & w & 6.4 & 7.0 & 0.0292 \\
            \hline
            \hline
            TSNAT-Small (34M) & w/o & 5.8 & 6.4 & 0.0054 \\
             \quad + Two Step Inference & w/o & 5.4 & 5.9 & 0.0173  \\
            \hline
            TSNAT-Middle (59M) & w/o & 5.4 & 6.0 & 0.0063 \\
             \quad + Two Step Inference & w/o & 5.2 & 5.7 & 0.0176  \\
            \hline
            TSNAT-Big (87M) & w/o & 5.3 & 6.0 & 0.0077 \\
              \quad + Two Step Inference & w/o & \textbf{5.1} & \textbf{5.6} & 0.0185  \\
            \hline
        \end{tabular}
        \begin{tablenotes}
            \item[$\lozenge$] These models additionally use speed-perturb to augment the speech data.
        \end{tablenotes}
        \vspace{-10pt}
    \end{threeparttable}
\end{table}

Compared with the other non-autoregressive models, we proposed TSNAT model can achieve quite competitive performance. To our best knowledge, the \texttt{TSNAT-Big} with two-step inference achieves the \textit{state-of-the-art} (SOTA) performance without depending on any external language models. The results also show that the two-step inference is faster than ST-NAT with a neural language model because the two-step inference of the dual-mode transformer decoder can be implemented in parallel without iteration step by step.

\subsubsection{The Analysis of Attention Weights in The Dual-Mode Decoder}
\begin{figure}[t]
    \centering
    \subfigure[Self-Attention of NAR Mode]{
        \centering
        \label{fig:slf_fst}
        \includegraphics[width=0.48\linewidth]{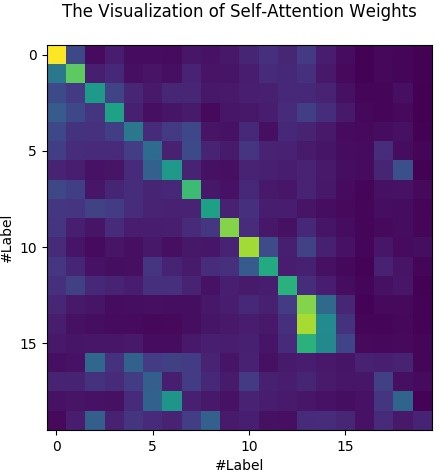}}
    \subfigure[Self-Attention of AR Mode]{
        \centering
        \label{fig:slf_sec}
        \includegraphics[width=0.48\linewidth]{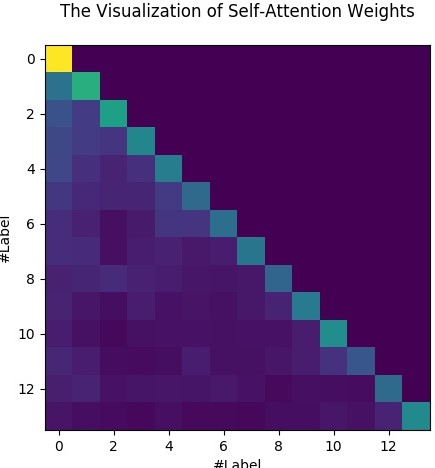}} \\
    \vspace{-10pt}
    \subfigure[EncDec-Attention of NAR Mode]{
        \centering
        \label{fig:src_fst}
        \includegraphics[width=\linewidth]{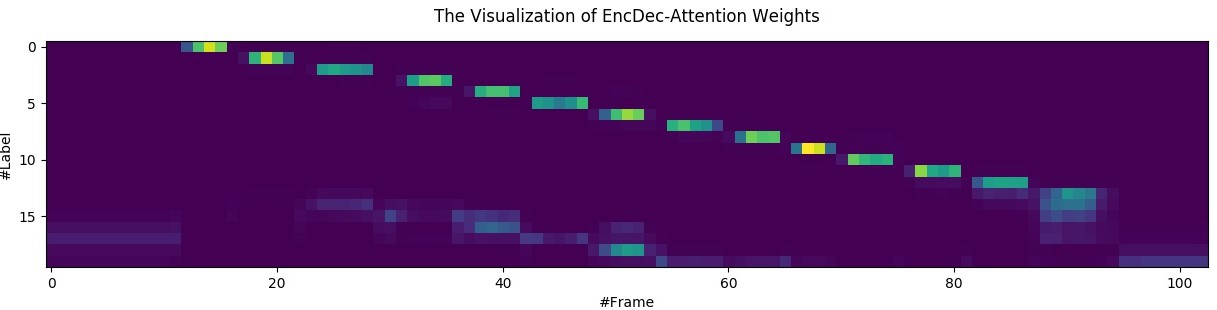}} \\
    \vspace{-10pt}
    \subfigure[EncDec-Attention of AR Mode]{
        \centering
        \label{fig:src_sec}
        \includegraphics[width=\linewidth]{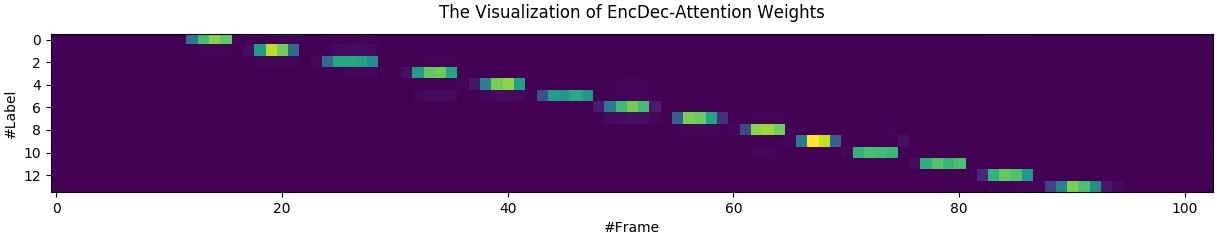}}
    \vspace{-10pt}
    \caption{The Visualization of Attention Weights in The Dual-Mode Decoder. We extract the self-attention weights and encoder-decoder attention weights from the last block of the dual-mode transformer decoder. The pictures come from the test set, and its corresponding sentence ID is 'BAC009S0764W0121'.}
    \vspace{-15pt}
\end{figure}
We want to find out the difference between the AR and NAR modes of the dual-mode transformer decoder. Therefore, we extract the attention weights of these two modes from the last decoder block. It's clear that the self-attention mechanism of the NAR model focuses on the whole input sequence, while the self-attention mechanism of the AR mode pays attention to the previous sequence. This is consistent with their own characteristics. Besides, the encoder-decoder attention weights of the NAR mode are more dispersed than the one under the AR mode, which also shows the autoregressive model has stronger modeling ability.
\vspace{-10pt}

\section{Conclusions}
Compared with the autoregressive model for speech recognition, the non-autoregressive model can get rid of the temporal dependency and predict the entire tokens in at least one step. However, the non-autoregressive model still faces two problems. On the one hand, there is still a great performance gap between the advanced autoregressive (AR) model and the non-autoregressive (NAR) transformer model. On the other hand, it's difficult for most of the non-autoregressive models to train and converge. In this paper, we try to address these problems from two aspects. Firstly, we propose a parameters-sharing training method to improve the performance of the NAR model by learning some valuable knowledge from the AR model. Then we make full use of the dual-mode decoder by introducing the two-step inference method. We conduct our experiments on a public Chinese mandarin dataset ASIEHLL-1. The results show that our proposed model can achieve comparable performance with the AR model and outperform other typical NAR models.

\bibliographystyle{IEEEtran}

\bibliography{mybib}

\end{document}